\documentclass[letter]{aa}  

\usepackage{graphicx}
\usepackage{txfonts}
\usepackage{natbib}
\bibpunct{(}{)}{;}{a}{}{,} 
\begin{document} 

\title{A candidate supermassive binary black hole system 
 in the brightest cluster galaxy of RBS~797}

\author{M. Gitti \inst{1,2,3}, M. Giroletti \inst{3}, G. Giovannini\inst{1,3},
L. Feretti\inst{3}, E. Liuzzo\inst{3}}

\institute{ 
Physics and Astronomy Department, University of Bologna, via Ranzani 1, 40127 Bologna, Italy \\ \email{myriam.gitti@oabo.inaf.it}
\and
INAF, Astronomical Observatory of Bologna, via Ranzani 1, 40127 Bologna, Italy\\
\and
INAF, Istituto di Radioastronomia di Bologna, via Gobetti 101, I-40129 Bologna, Italy }

\authorrunning{Gitti et al.} 
\titlerunning{A candidate supermassive binary black hole system in the brightest cluster galaxy of RBS~797}

\date{Accepted 14 August 2013}

\abstract 
{
  The radio source at the center of the cool-core galaxy cluster RBS
  797 (z=0.35) is known to exhibit a misalignment of its radio jets
  and lobes observed at different VLA scales, with the innermost
  $\sim$ kpc-scale jets being almost orthogonal to the radio emission
  which extends for tens of kpc filling the X-ray cavities.  Gitti et
  al. suggested that this peculiar radio morphology may indicate a
  recurrent activity of the central radio source, where the jet
  orientation is changing between the different outbursts due to the
  effects of supermassive binary black holes (SMBBHs).}
{
  We aim to reveal the nuclear radio properties of the brightest
  cluster galaxy (BCG) in RBS~797 and to investigate the presence of a
  SMBBH system in its center. }
{
  We performed new high-resolution observations at 5 GHz with the
  European VLBI Network (EVN) on May 3, 2013, reaching an angular
  resolution of $\sim 9 \times 5$ mas$^2$ and a sensitivity of 36
  $\mu$Jy/beam. We also re-analyzed VLA archival data at 4.8
  GHz in A- and B- configurations.}
{
  We report the EVN detection of two compact components in the BCG of
  RBS~797, with a projected separation of $\sim$77 pc.  We can
  envisage two possible scenarios: the two components are two
  different nuclei in a close binary system, or they are the core and
  a knot of its jet.  Both interpretations are consistent with the
  presence of SMBBHs.  Our re-analysis of VLA archival data seems to
  favor the first scenario, as we detect two pairs of radio jets
  misaligned by $\sim 90$ degrees on the same $\sim$ kpc scale
  emanating from the central radio core.  If the two outbursts are
  almost contemporaneous, this is clear evidence of the presence of
  two active supermassive black holes whose radio nuclei are
  unresolved at VLA resolution.  The nature of the double source
  detected by our EVN observations in the BCG of RBS~797 can be
  established only by future sensitive, multi-frequency VLBI
  observations.  If confirmed, RBS~797 would be the first SMBBH system
  observed at medium-high redshift at VLBI resolution. } {}


\maketitle

\vspace{-0.4in}
\section{Introduction} 
\vspace{-0.1in}

In the standard cold dark matter cosmological scenario, the formation
of structures is a hierarchical process acting through galaxy mergers
to form larger and larger structures, up to clusters of galaxies
\citep[e.g.,][]{White-Rees_1978}.  During the merging of galaxies, the
supermassive black holes (SMBHs) at their centers form binary systems,
most of which, it is believed, eventually merge in less than a Hubble
time \citep[e.g.,][]{Komossa_2003}.  An understanding of how SMBHs
form and coalesce is important for the understanding of active
galactic nuclei (AGN) dynamics as well as galaxy formation in general.
However, observational cases where both SMBHs in a merging system are
accreting as AGNs are rare, and there have only been a few confirmed
kpc-scale binary AGNs detected via various techniques
\citep[e.g.,][where in the last case the separation is $\sim$150
pc]{Komossa_2003, Hudson_2006, Koss_2011, Fabbiano_2011}.  Recent
simulations show that selection of double-peaked narrow emission lines
is a promising method for identifying dual SMBH candidates, but
high-resolution, multi-wavelength follow-up observations are of
critical importance for confirming the nature of the candidates
\citep[][and references therein]{Blecha_2013}.  On the other hand,
more compact binaries (with separations of less than few hundred pc)
are extremely difficult to resolve with present telescopes, and only
one object has been confirmed to host a pc-scale supermassive binary
black hole (SMBBH) system \citep{Rodriguez_2006, Burke_2011}.
\vspace{-0.05in}

The X-ray luminous galaxy cluster RBS~797 (z=0.35) was discovered in
the {\it ROSAT} All-Sky Survey \citep[RASS,][]{Schwope_2000} and then
observed by the optical ROSAT Bright Survey (RBS) during the
identification of the X-ray RASS sources \citep{Fischer_1998,
  Schwope_2000}.  Follow-up observations with {\it Chandra} revealed
that it is a cool-core cluster, showing two symmetric X-ray minima
(the so-called X-ray {\it cavities}) with diameters $\sim$20 kpc and a
bright X-ray point-like source in the cluster center (Schindler et
al. 2001).  Comparison of the {\it HST} and {\it Chandra} images
reveals that the central brightest cluster galaxy (BCG) that shows
optical emission lines typical of AGNs \citep{Fischer_1998} coincides
with the nuclear X-ray source.  The optical BCG has the appearance of
being bifurcated, perhaps because of an absorbing dust lane
\citep{Cavagnolo_2011}.
In the NRAO VLA Sky Survey (NVSS), an unresolved radio source of 20
mJy is present at the center of RBS~797. The discovery of the X-ray
cavity system has motivated in the past decade new radio observations
at different frequencies and angular resolutions to study in detail
the interaction of the central radio source with the intra-cluster
medium (ICM).  In particular, \citet{Gitti_2006} performed VLA radio
observations of RBS~797 at 1.4 and 4.8 GHz, ranging in resolution from
a few tens of arcsec to subarcsec, reporting the detection of radio
emission on different scales and orientations. The 1.4 GHz radio
emission imaged at $\sim$1 arcsec resolution is found by these authors
to fill the X-ray cavities nicely, extending along the
northeast-southwest direction up to $\sim 43$ kpc from the cluster
center. This is the typical configuration observed in most cool-core
clusters: the radio jets from the cluster central elliptical extend
outwards in a bipolar flow, inflating lobes of radio-emitting plasma
(radio {\it bubbles}) which push aside the X-ray emitting gas thus
excavating depressions in the ICM which are detectable as apparent
cavities in the X-ray images \citep[see, e.g.,][for recent reviews on
AGN feedback in galaxy clusters]{Gitti_2012, McNamara-Nulsen_2012}.
Remarkably, the innermost 4.8 GHz radio jets imaged at subarcsec
resolution are not aligned with the direction of the radio and X-ray
emission seen at larger scale, but extend out from the core up to
$\sim 13$ kpc and clearly show a north-south orientation, they are,
therefore, almost {\it perpendicular} to the axis of the 1.4 GHz
emission filling the X-ray cavities \citep[see Fig. 4
of][]{Gitti_2006}.  In-depth studies of the X-ray cavities and radio
emission in RBS~797 have been presented by \citet{Gitti_2006},
\citet{Cavagnolo_2011}, and \citet{Doria_2012}. The cluster RBS~797
was also included in the cavity sample studies of \citet{Birzan_2004},
\citet{Birzan_2008}, \citet{Rafferty_2006}, and
\citet{Hlavacek_2012}. However, no definite explanation for the
observed change in radio orientation has been presented so far.
We note that according to the existing radio images the central source
does not show a slowly precessing jet, but a real change in the radio
emission position angle (P.A.).  Because this galaxy is the cluster
BCG, we do not expect a large influence by galaxy motion.  As already
suggested by \citet{Gitti_2006}, we argue that this peculiar radio
structure could be explained in the context of SMBBH models.  The
presence of a SMBBH system in a galactic nucleus may indeed become
manifest through a change in the jet P.A., which is likely originated
by a spin flip of the primary SMBH caused perhaps by capture of a
second SMBH, or to black hole mergers \citep{Begelman_1980,
  Merritt_2005}.  Hence, relativistic material ejected in different
episodes of activity may be expelled in different directions from the
central engine, leaving a fossil record of the orientation history of
the jets in the radio lobes.  The radical change in the jet
orientation may also be only apparent, being in fact due to the
presence of {\it two pairs of radio jets} ejected in different
directions by two close SMBHs which are both active.
In all these scenarios, the different radio directions observed in
RBS~797 may thus be due to the presence of SMBBHs in the center of the
BCG, unresolved at VLA resolution.

In this paper we present new high-resolution European Very Long
Baseline Interferometry (VLBI) Network data of the radio source in the
BCG of RBS~797 to study its nuclear properties and to investigate the
presence of a SMBBH system in its center.
We also show new Very Large Array (VLA) images at (sub-)arcsec
resolutions produced by our re-analysis of combined archival data.
With $H_0 = 70 \mbox{ km s}^{-1} \mbox{ Mpc}^{-1}$, and
$\Omega_M=1-\Omega_{\Lambda}=0.3$, the luminosity distance of RBS~797
is 1858 Mpc and $1 ''$ corresponds to $\sim$4.8 kpc.

\vspace{-0.2in}
\section{Observations and Data Reduction}
\vspace{-0.05in}

We observed the radio core of the BCG in RBS~797 on May 3, 2013
(project code: RSG05, PI: M. Gitti) with a subset of the European VLBI
Network (EVN) including the stations of Effelsberg, Medicina, Noto,
Onsala, Shanghai, Torun, and Yebes. We observed at 5 GHz using the
e-VLBI technique, with data acquired and transmitted in real-time from
the stations to the EVN data processor at JIVE. A bandwidth of 1 Gbps
was sustained by most stations, corresponding to 128 MHz divided into
eight sub-bands, with two polarizations and 2-bit sampling.

We observed in phase reference mode, using the nearby ($d=1.85^\circ$)
source J0954+7435 as a phase calibrator, with 14 $\times$ 4-minute
scans on the target bracketed by 60\,s scans on the calibrator, for a
total net on source time of about 56 minutes. We also observed four
nearby NVSS sources (J094729+764355, J093501+754526, J094045+764833,
and J095143+752952) to explore them as candidate phase reference
sources for future observations; none of them turned out to host a
compact component suitable for this purpose.

Correlation was performed in real time at JIVE; the JIVE pipeline was
also used to carry out a priori amplitude calibration, automated
flagging, and fringe fitting with AIPS (Astronomical Image Processing
System) tasks. We edited the final visibility data and produced clean
images in Difmap. Owing to the low flux density in the source, we did
not attempt self-calibration.

The final beam size with natural weights is $9.4 \times 5.3$ mas$^{2}$
in P.A.\ $-24^\circ$. The image noise r.m.s.\ is $\sim 36\,
\mu$Jy/beam, i.e., consistent with the expected thermal noise.
Systematic amplitude calibration errors are typically at 5\%;
therefore, we assume this uncertainty on the flux density
measurements.

\vspace{-0.15in}
\section{Results}
\label{results.sec}
\vspace{-0.05in}

\begin{figure}[t]
\hspace{0.3in}
\includegraphics[width=2.8in]{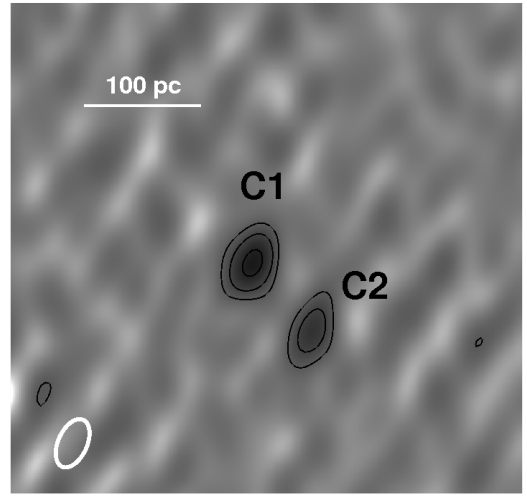}
\vspace{-0.1in}
\caption{\label{evn.fig} \small 5 GHz EVN map of the BCG in RBS~797 at a
  resolution of $9.4 \times 5.3$ mas$^2$ in P.A.\ $-24^\circ$ (the
  beam is shown in the lower-left corner). The r.m.s. noise is 36
  $\mu$Jy/beam and the peak flux density is 0.53 mJy/beam. The
  contours levels start at 3$\sigma$ and increase by a factor of 2.
  Two components, separated by 16 mas ($\sim$ 77 pc), are clearly
  detected. }
\vspace{-0.1in}
\end{figure}

\begin{table}
\begin{center}
  \caption{ Gaussian Model Components.  }
\vspace{-0.2in}
\begin{tabular}{lccccccc}
\hline
\hline
Comp.  & Flux  & r      & $\theta$ & a & a     & b/a &  $\Phi$  \\
~  & (mJy) & (mas)  & (deg)    & (mas) & (pc)  &    & (deg) \\
\hline
~&~&~&~&~&~&~&~\\
{\bf C1} & 0.606 & 0.43   & 29.4 & 2.96  & 14.2    &    1.00&     58.1  \\
{\bf C2} & 0.540 & 16.1  &-139.2& 7.44   &  35.7   &    0.59&   -1.4    \\
\hline
\label{modelfit.tab}
\end{tabular}
\vspace{-0.15in}
\tablefoot{\small
  Col. (1): Gaussian component as labeled in
  Fig. \ref{evn.fig}. Col. (2): Flux density. Cols. (3)-(4): Polar
  coordinates of the center of the component relative to the
  observation pointing RA: 09$^{\rm h}$ 47$^{\rm m}$ 12$^{\rm s}$.760,
  Dec: $+76^{\circ}$ 23$'$ 13$''$.740.  Cols. (5)-(6): Major axis.
  Col. (7): Axial ratio.  Col. (8): Component orientation. All angles
  are measured from north to east.
}
\end{center}
\vspace{-0.2in}
\end{table}

Our 5 GHz EVN map of the BCG in RBS~797 is shown in
Fig. \ref{evn.fig}.  Two components are clearly detected. The first
component, labeled C1, is located at RA: 09$^{\rm h}$ 47$^{\rm m}$
12$^{\rm s}$.760, Dec: $+76^{\circ}$ 23$'$ 13$''$.733 and has a total
flux density of 0.61 mJy.  The second component, labeled C2, is
separated by 16 mas ($76.8$ pc) in P.A. −139$^{\circ}$ and has a total
flux density of 0.54 mJy.  The results of the visibility model fit
with two Gaussian components (Table \ref{modelfit.tab}) show that both
components are compact and smaller than the observing beam.  On the
other hand, the size estimates provided by the model fit indicate that
component C2 appears less compact than component C1 along the major
axis, although it can still be considered unresolved given the size
and shape of the beam.

With the aim of getting more insights into the global properties of
the radio source, we also re-analyzed the archival VLA A- and B-array
data at 4.8 GHz.  To fully exploit the relative advantages in terms of
angular resolution and sensitivity of these two VLA configurations, we
merged (with the AIPS task {\ttfamily DBCON}) the two dataset to
produce a new combined A$+$B dataset at 4.8 GHz.  We then produced
images at different resolutions by specifying appropriate values of
the parameters {\ttfamily UVTAPER} and {\ttfamily ROBUST} in the task
{\ttfamily IMAGR}.  Our new 4.8 GHz VLA maps are shown in
Fig. \ref{chandra-vla.fig}, superposed on the {\it Chandra} image of
the central cluster region.

The subarcsec resolution image (black contours in
Fig. \ref{chandra-vla.fig}, zoomed in the right panel) cleary shows
the presence of another jet-like feature to the east, not detected in
the previously published maps, which is the likely counterpart of the
one to the west.  In addition to the well-known nort-south jets, we
therefore find strong evidence of the presence of 4.8 GHz jets
emanating also to the east-west direction. This east-west emission is
best visible in the arcsec resolution image (green contours in
Fig. \ref{chandra-vla.fig}); it extends up to $\sim$ 35 kpc from the
center and is clearly shown for the first time to nicely fill the
X-ray cavity also at 4.8 GHz, besides the already known 1.4 GHz
emission \citep[cf. Fig. 4 of][]{Gitti_2006}.

\vspace{-0.15in}
\section{Discussion}

\vspace{-0.05in}
\subsection{Nature of the VLBI double source}
\label{disc-evn.sec}
\vspace{-0.05in}

The detection of two compact components at VLBI resolution at the
center of the BCG in RBS~797 is remarkable, especially when the short
time of our observation is considered (only 56 minutes on source
obtained over a limited hour angle range).
By comparison, in a complete sample of 34 BCGs in nearby Abell
clusters \citep[distance class $<$3,][]{Liuzzo_2010} the 5 GHz VLBI
detection rate is about $68\%$, with only one case of double source.
We note that this source, 3C 75 in A 400, is a dumbbell galaxy where
the two components are separated by a much larger distance ($\sim$7.2
kpc in projection) than those detected in RBS~797.

The interpretation of the origin and nature of the double source in
RBS~797 is complex, and with the current data we can envisage two
different scenarios:
(1) the two components C1 and C2 are two different nuclei in a close
binary system, or
(2) the two components are the core and a knot of its jet, and the
emission from the underlying jet flow connecting the two components is
not visible because of the limited sensitivity of our short
observation.
The presence of SMBBHs would be obvious in the first scenario, but
given the VLA-scale properties of RBS~797 it would also be likely in
the second scenario (see Sect. \ref{disc-vla.sec}).

In the first scenario, we expect sensitive, multi-frequency VLBI
observations to measure a flat spectrum for both components C1 and C2,
and to reveal at least one jet emanating from one of them. If the two
nuclear components are both active, each of them may show its own jet
(or jet-counterjet pair) emerging from its center. We stress that only
two cases, both at low redshift ($z < 0.06)$, are known in the
literature, namely 3C 75 in A 400 \citep{Owen_1985} and 0402+379
\citep{Rodriguez_2006}, which is the only pc-scale SMBBH system
discovered so far by the VLBI \citep{Burke_2011}.  The detection of
dual compact radio (or X-ray) sources in an active galaxy provides the
most unambiguous evidence that a system hosts SMBBHs. Therefore, if
the two compact radio components separated by only $\sim$77 pc are
confirmed to be two nuclei, RBS~797 would represent the {\it first}
case of active SMBBHs observed at medium-high redshift at VLBI
resolution.  Considering that the sub-kpc-scale dual AGN may be
relatively rare \citep[as suggested by the recent observational work
of][]{Comerford_2012}, this detection would be particularly notable.

In the second scenario, component C1 is the most likely main core
candidate because of its more compact size and the higher flux density
(brightness temperature of $\sim 5 \times 10^6$ K); we also estimate
that C2 has a brightness temperature of $\sim 10^6$ K, which is a
reasonable value for a typical jet component.
However, it is still possible that a knot is brighter than the nucleus
itself, so the role of the nucleus-jet components in our
interpretation may be inverted.
A final test for this scenario can only be carried out by
investigating the spectral properties of the two components, revealing
a flat-spectrum core and a steep-spectrum jet knot. Deep observations
may also reveal the presence of radio emission connecting the two
features.  However, we note that the orientation of the C1-C2 vector
does not match any of the two P.A. of the large-scale radio emission
(see Sect. \ref{disc-vla.sec}).

With the current EVN data it is not possible to reach a definite
conclusion about the origin and nature of the two compact components
detected in the BCG of RBS~797, and their interpretation remains an
open issue.  However, from the morphology and orientation of the radio
emission observed at VLA-scale we can gather elements in favor of one
scenario rather than the other.

\vspace{-0.15in}
\subsection{Hints from archival VLA observations} 
\label{disc-vla.sec}
\vspace{-0.05in}

\begin{figure}[t]
\includegraphics[width=3.5in]{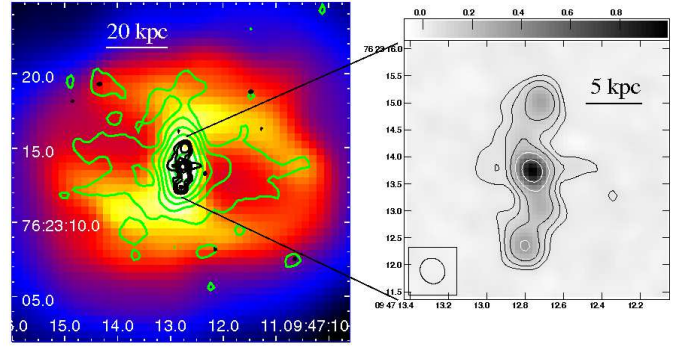}
\vspace{-0.1in} 
\caption{\label{chandra-vla.fig} \small The 4.8 GHz VLA contours obtained
  from the combined A- and B-array archival observations of RBS~797,
  imaged at different resolutions, are overlaid onto the
  \textit{Chandra} image of the central region of the cluster (left
  panel). 
  {\it Green contours}: 4.8 GHz VLA map at $1''.38 \times 1''.33$
  resolution, obtained by setting {\ttfamily ROBUST=+5, UVTAPER=250};
  the r.m.s. noise is 0.01 mJy/beam and the contour levels are 0.03,
  0.06, 0.12, 0.24, 0.48, and 0.96 mJy/beam; the total flux density is
  $\sim 4$ mJy, with a peak flux density of $1.5$ mJy/beam.  {\it
    Black contours} (best visible in the zoom in the right panel): 4.8
  GHz VLA map at $0''.49 \times 0''.44$ resolution, obtained by
  setting {\ttfamily ROBUST=0, UVTAPER=0}; the r.m.s. noise is 0.01
  mJy/beam and the contour levels are 0.04, 0.08, 0.16, 0.32,
  and 0.64 mJy/beam; the total flux density is $\sim 2.8$ mJy,
  with a peak flux density of $1.0$ mJy/beam.  }
\vspace{-0.1in}
\end{figure}

As first discovered by \citet{Gitti_2006} and then confirmed by
\citet{Birzan_2008} and \citet{Cavagnolo_2011}, the central radio
source in RBS~797 observed at VLA resolutions shows different
orientations of the radio jets and lobes with scale.  In particular,
as seen in projection, the innermost 4.8 GHz jets which extend to the
north-south direction up to $\sim 13$ kpc from the radio core are
almost orthogonal to the axis of the 1.4 GHz radio emission filling
the X-ray cavities and extending to the northeast-southwest direction
up to $\sim 43$ kpc from the core.  As a possible interpretation of
this peculiar morphology, \citet{Gitti_2006} suggested that the
central radio source is experiencing recurrent activity, with
misaligned jet axes from its different episodes, which could point to
the effects of SMBBHs. These authors also noted the presence of a
feature extending to the west out to a distance of $\sim 9$ kpc from
the 4.8 GHz core (see their Fig. 2c) that, if real, could be related
to the 1.4 GHz emission filling the cavities.
We stress that, if confirmed, this western jet-like feature on $\sim$
kpc scale would further strengthen the interpretation that RBS~797
harbors SMBBHs, as it may indicate the presence of ejecta originating
from the secondary (active) SMBH.

Our re-analysis of the archival VLA A- and B-array data at 4.8 GHz
confirms that the western 4.8 GHz jet structure is real and not an
image artifact, indicating the presence of kpc-scale jets also
emanating to the east-west direction (see black contours in
Fig. \ref{chandra-vla.fig}).
In the following discussion, we note that the 5 GHz EVN map in
Fig. \ref{evn.fig} spans $\sim 80$ mas ($\sim 400$ pc), corresponding
to a small region in the central part of the 5 GHz VLA image in
Fig. \ref{chandra-vla.fig}, which shows structures on scales of $\sim$
20 kpc.

Since the large-scale east-west emission aligned with the X-ray cavity
axis appears to be more extended and diffuse than the north-south one
(see green contours in Fig. \ref{chandra-vla.fig}), it should be
older. This would be consistent with the fact that it has already
created two large X-ray cavities, whereas the north-south emission is
likely to be still in the process of excavating the ICM.  In this
context, \citet{Cavagnolo_2011} report a hint of barely resolved
structure in the {\it Chandra} image associated with the 4.8 GHz radio
structure, with residual X-ray deficits just beyond the tips of the
innermost radio jets, which might suggest potentially small cavities
from a new outburst episode.  This scenario would be consistent with
the presence of a SMBBH system in which the observed radio emissions,
elongated in different directions, are due to recurrent activity from
a single central object that has changed ejection orientation because
of the interaction with the secondary SMBH \citep[due, for example, to
merger or spin-flip, e.g.,][and references therein]{Merritt_2005}.
The coalescent binaries able to alter the spin axis of the primary
SMBH typically stall at a separation \citep[$\sim$1 pc,][and
references therein]{Merritt-Ekers_2002} that is smaller than the
observed separation ($\sim$77 pc) of the two compact radio components
in RBS 797. Therefore, in this picture the VLBI double source would
probably be the core-jet structure of the primary SMBH (second
scenario discussed in Sect. \ref{disc-evn.sec}), whereas the secondary
SMBH causing the spin-flip of the primary would remain undetected.

On the other hand, with the new combined A$+$B 4.8 GHz VLA images
shown in Fig. \ref{chandra-vla.fig} (black contours), we find a
different orientation of the radio jets {\it on the same $\sim$kpc
  scale}, indicating that the radio-emitting plasma on the east-west
direction is still being freshly injected (in the absence of
multi-frequency data needed for a detailed spectral aging analysis, we
estimated a kinematic age of $\approx 10^5$ yrs by assuming a typical
jet velocity of $\sim$0.1 c).  The two outbursts could thus be almost
contemporaneous, indicating the presence of two active SMBHs.  This
would support the first scenario discussed in
Sect. \ref{disc-evn.sec}.
We also note that in the light of the VLA-scale properties, the
core-jet hypothesis (the second scenario discussed in
Sect. \ref{disc-evn.sec}) is challenging since the pc-scale jet flow
would not be aligned with any of the directions seen at kpc-scale in
the VLA images.

\vspace{-0.15in}
\section{Conclusions}
\vspace{-0.05in}

Our main results can be summarized as follows:
\vspace{-0.1in}
\begin{itemize}

\item We report the VLBI detection of two compact radio components
  separated by $\sim$77 pc in the BCG of the cool-core cluster
  RBS~797, obtained by our new EVN observations performed on May 3,
  2013 at 5 GHz.

\item Our re-analysis of the archival 4.8 GHz VLA data shows strong
  evidence of the presence of radio jets at subarcsec resolution also
  emanating to the east-west direction, in addition to the already
  known north-south jets.  We therefore find different orientations of
  the jets on the same $\sim$kpc scale, indicating the likely presence
  of contemporaneous emission from two active SMBHs.

\item We discuss the possible scenarios for the nature and origin of
  the VLBI double source, which suggest the presence of a SMBBH
  system.  However, without detailed information on the morphological
  and spectral properties of the radio emission it is not possible to
  establish if the second component is in fact a different nucleus, or
  if it is a knot within the jet of the first nuclear component. In
  particular, in the first case we would expect to measure a flat
  spectrum for both components and to reveal jets emerging from each
  of them if they are both active.
\end{itemize}
\vspace{-0.1in}
New sensitive, high-resolution, multifrequency VLBI observations are
essential to unveil the nuclear radio properties of the BCG in
RBS~797, and represent the only way to confirm the presence of SMBBHs
in its center.

\vspace{-0.15in}
\section*{Acknowledgments}
\vspace{-0.05in}
We thank the referee for constructive comments that improved the
presentation of the work.  We acknowledge the financial contribution
from contract ASI - I/009/10/0.  The European VLBI Network
(http://www.evlbi.org/) is a joint facility of European, Chinese,
South African, and other radio astronomy institutes funded by their
national research councils.

\vspace{-0.15in}
{\small
\bibliographystyle{aa}
\bibliography{bibliography-gitti}
}

\end{document}